\documentclass[runningheads]{llncs}
\usepackage[T1]{fontenc}
\usepackage{graphicx}
\usepackage{hyperref}
\usepackage{color}
\usepackage{algorithm}
\usepackage{algpseudocode}
\usepackage{multirow}
\usepackage{wrapfig}
\usepackage{todonotes}
\usepackage{amssymb}
\usepackage{array}
\usepackage{enumitem}
\usepackage{longtable}
\usepackage{tabularx}
\usepackage{comment}
\usepackage{amsmath}

\usepackage{tikz}

\newenvironment{tinylongtable}
 {\tiny\longtable}
 {\endlongtable}

\newcommand{\mypar}[1]{\smallskip\noindent\textbf{#1.}}
\newcommand{\rom}[1]{%
  \textup{\uppercase\expandafter{\romannumeral#1}}%
}

\begin{document}
\title{Design of a Quality Management System based on the EU Artificial Intelligence Act}

\titlerunning{QMS based on the EU AIA}
%

\author{Henryk Mustroph \and Stefanie Rinderle-Ma}
\authorrunning{H. Mustroph and S. Rinderle-Ma}
%
%

\institute{Technical University of Munich, TUM School of Computation, Information and Technology, Garching, Germany\\
\email{\{henryk.mustroph,stefanie.rinderle-ma\}@tum.de}}

\maketitle

\begin{abstract}
The EU AI Act mandates that providers and deployers of high-risk AI systems establish a quality management system (QMS). Among other criteria, a QMS shall help verify and document the AI system design and quality and monitor the proper implementation of all high-risk AI system requirements. Current research rarely explores practical solutions for implementing the EU AI Act. Instead, it tends to focus on theoretical concepts. As a result, more attention must be paid to tools that help humans actively check and document AI systems and orchestrate the implementation of all high-risk AI system requirements. Therefore, this paper introduces a new design concept and prototype for a QMS as a microservice Software as a Service web application. It connects directly to the AI system for verification and documentation and enables the orchestration and integration of various sub-services, which can be individually designed, each tailored to specific high-risk AI system requirements. The first version of the prototype connects to the Phi-3-mini-128k-instruct LLM as an example of an AI system and integrates a risk management system and a data management system. The prototype is evaluated through a qualitative assessment of the implemented requirements, a GPU memory and performance analysis, and an evaluation with IT, AI, and legal experts.


\end{abstract}

%
%
%
%
\section{Introduction}
\label{sec:introduction}
\begin{figure}[!ht]
    \centering
    \includegraphics[width=1.0\linewidth]{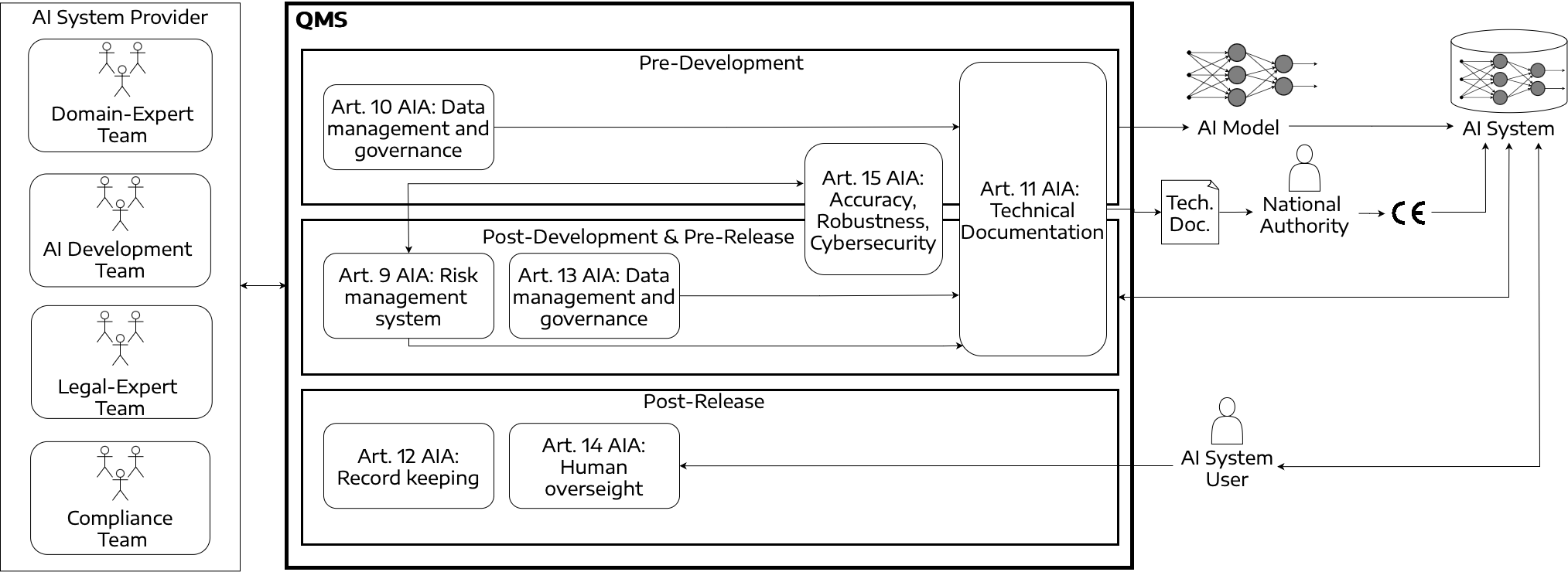}
    \caption{Proposed design concept for a QMS based on the EU AI Act}
    \label{fig:qmsAIA_idea}
\end{figure}

Over recent years, the rise of Artificial Intelligence (AI) has introduced rapid advancements and heightened risks, particularly in critical sectors like medicine, finance, and law, where AI increasingly participates in or even controls decision-making processes. To have risks under control, the European Union (EU) Commission introduced the EU AI Act (AIA) in 2021, which came into legal force on 1 August 2024. Since 2021, researchers from the legal tech and information systems community have been working on standardized and unified implementations of the AIA. Many are mainly concerned with developing theoretical ideas, processes, and frameworks on how AI compliance management can be organized in a standardized and organized way (cf. \cite{clarke2019_bpm_for_resp_ai,mökander2023_auditing_llm,ortega2023_ai_product_lifecycle,novelli2023_bpm_aia}). However, more research is needed on practical tools for efficiently implementing the AIA and involving humans in the AI compliance management process. For instance, a Quality Management System (QMS) responsible for verifying compliance is essential for demonstrating adherence to AIA regulations, particularly for high-risk AI systems. Therefore, this technical report proposes and explains a design concept for an all-in-one QMS solution illustrated in Figure \ref{fig:qmsAIA_idea}.

The goal is to offer a QMS as a Software as a Service (SaaS) web application, which includes several sub-services that function as independent systems. These sub-services are organized into three core phases of the AI system lifecycle:  \emph{pre-development}, \emph{post-development \& pre-release}, and \emph{post-release}. Each sub-service can be individually designed to implement specific requirements of the AIA for high-risk AI systems. The QMS shall assist AI system providers in managing compliance, verification, and documentation processes for AI systems. It shall connect with the AI systems to perform technical assessments directly within the QMS. The sub-services shall interact with each other, allowing data to be transferred between them. For example, data from the sub-service addressing Art. 9 AIA can be incorporated into technical documentation, which can then be submitted to national authorities as proof of compliance. Such a QMS can enhance the traceability and reliability of AI compliance management, making it more transparent and allowing for continuous improvement through process mining techniques, as recommended by \cite{Pery2021_processmining_aia}.

This technical report outlines the legal and structural requirements based on the AIA. It presents a conceptual architecture and data model for a QMS aligned with these requirements. It also introduces a first-version prototype QMS, which integrates and adapts a Large Language Model (LLM) as an example of an AI system. The paper presents a summary of the AIA and related work Sect. \ref{sect:pre}. The method in Sect. \ref{sec:method} outlines the requirements, the QMS architecture, and details of the implemented prototype. The prototype is evaluated and discussed in Sect. \ref{sec:evaluation}. The conclusion is provided in Sect. \ref{sec:disc}.

%
%
%
%
\section{EU Artificial Intelligence Act \& Related Work}
\label{sect:pre}
A summary of the EU AI Act (AIA) is given, followed by related work dealing with the AIA.

\subsection{EU Artificial Intelligence Act}
\label{sect:pre:aia}
The AIA \cite{EU_AIA2024} is a comprehensive legislative framework regulating the development, deployment, and use of AI systems within the European Union. According to the AIA, an AI system \emph{``means a machine-based system that is designed to operate with varying levels of autonomy and that may exhibit adaptiveness after deployment, and that, for explicit or implicit objectives, infers, from the input it receives, how to generate outputs such as predictions, content, recommendations, or decisions that can influence physical or virtual environments.''} \footnote{Chapter 1 Article 3 para. 1 AIA}.
The AIA categorizes AI systems into four different risk classes. Depending on which risk category the AI system is in, specific requirements must be met to be authorized in the EU. The following AI system risk classes exist in the AIA:
\begin{itemize}
  \item \textbf{Unacceptable Risk}: AI practices pose unacceptable safety, rights, and societal well-being risks. One example would be a social credit system, which evaluates human behavior and actions. These AI systems will be prohibited in the EU. \footnote{Chapter 2 Article 5 AIA}
  \item \textbf{High Risk}: These AI systems are allowed in the EU, but can cause risks to critical areas such as health, safety, and fundamental rights and therefore need to comply with a couple of regulations. \footnote{Chapter 3 Section 1 Article 6 AIA}
  \item \textbf{Limited or Low Risk:} AI systems with limited risks have limited transparency regulations. An example of such an AI system is a chatbot answering customer requests employed in a non-high-risk domain. Low-risk AI systems such as spam filters, have no transparency regulations.
\end{itemize}

High-risk AI systems must adhere to the most challenging regulations and furnish the EU with documented evidence detailing risk identification, analysis, assessment, mitigation, data management, and data governance. They must be evaluated through their whole lifecycle before entering and post-market. Furthermore, they must demonstrate through technical evaluation that despite being classified as high-risk applications, measures are in place to limit risks effectively. Additionally, GPAI models present a unique challenge in classification under the risk categories. They are versatile and can be employed across various tasks, domains, and purposes, as mentioned by \cite{novelli2023_ai_risks}. For that, the FLI came up in 2022 with an article that presents a short list of recommendations that suggested treating GPAI systems the same as high-risk AI systems \cite{fli2022_gpai}. They suggested establishing a QMS for GPAIs, the same as for high-risk AI systems, and checking as many regulations as possible from Chapter 3, Section 2. In \cite{ortega2023_ai_product_lifecycle}, the authors also plead for categorizing GPAI systems as high-risk AI systems. The AIA partly adopted the recommendations from the FLI. It defined a GPAI model as a model \emph{``... trained with a large amount of data using self-supervision at scale, that displays significant generality and is capable of competently performing a wide range of distinct tasks regardless of the way the model is placed on the market...''} \footnote{Chapter 1 Article 3 para. 63 AIA}. Based on that definition, GPAI models can also be used in high-risk domains, for high-risk tasks, or integrated into high-risk systems. However, for GPAI models alone, no QMS system as for high-risk AI systems is mandatory; only transparency of used training, validation, and testing data, potential drawbacks, and risk should be documented, which is less strict but could still be changed in future adoptions of the AIA.
Additionally, the AIA defined GPAI models with systematic risk, which is, \emph{``... a risk that is specific to the high-impact capabilities of general-purpose AI models, having a significant impact on the Union market due to their reach, or due to actual or reasonably foreseeable negative effects on public health, safety, public security, fundamental rights, or the society as a whole...''} \footnote{Chapter 1 Article 3 para. 65 AIA}. GPAI models with systematic risks are those that use \(10^{25}\) FLOPS for training, such as GPT-4, which used \(\sim 2*10^{25}\) FLOPS with 1.76 trillion parameters \footnote{Chapter 5 Section 3 Article 55 AIA}. These models must perform more evaluations and risk assessments and adhere to transparency and documentation obligations. Because most open-source GPAI models are much smaller and trained with less data, they will probably not be treated as having systematic risk.

\mypar{High-Risk AI Systems Regulations} A Summary of the key regulations for high-risk AI systems \footnote{Chapter 3 Section 2 AIA} is given. For detailed provisions, refer to the complete articles in the AIA \cite{EU_AIA2024}.

\begin{itemize}
  \item \textbf{Article 9 Risk Management System} The article states that a Risk Management System (RMS) should be ``established, applied, documented and maintained'' (para. 1). The article describes the RMS as an \emph{iterative process} that shall be documented and updated throughout the entire life cycle of the high-risk AI system (para. 2). In addition, the process is specified in the letters. Foreseeable risks shall be identified, analyzed, evaluated, and mitigated (lit. a). Risks that could result in a possible misuse of the AI system shall also be documented and mitigated (lit. b). Risks after placing the system on the market shall be identifiable (lit. c). Targeted risk management measures shall be applied (lit. d). Risk management measures shall be taken in such a way that as few interactions as possible occur (para. 4) but also that residual risks are still considered acceptable (para. 5). To identify risks, the underlying tests shall be carried out under real-world conditions before, during and after the development phase and post-market (para. 6, 7, 8).
  
  \item \textbf{Article 10 Data and data governance} For data used to train high-risk AI systems, it shall be ensured that the training, validation, and testing dataset splits meet the specified quality criteria in para. 2 ff. (para. 1). These datasets shall be governed and managed according to practices suited to the AI system’s intended purpose (para. 2). Key aspects include design choices, data collection and preparation, data suitability, bias examination and mitigation, and data gaps (para. 2 lit. a - h). Training, validation, and testing datasets shall be relevant, representative, to the best extent, free of errors, and as complete as possible for their intended purpose (para. 3).
  
  \item \textbf{Article 11 Technical documentation} The technical documentation shall be prepared and submitted before the high-risk AI system is placed on the market and shall be continuously updated (para. 1). It shall contain evidence for at least all points listed in Annex \rom{4} AIA, including all regulations for high-risk AI systems (para. 2).
  
  \item \textbf{Article 12 Record-keeping} High-risk AI systems shall be designed to automatically record event logs throughout the AI system’s entire lifetime (para. 1).

  \item \textbf{Article 13 Transparency and provision of information to deployers} High-risk AI systems shall be designed to provide sufficient transparency, allowing deployers to interpret and use the AI system’s output correctly (para 1). Additionally, high-risk AI systems shall come with user instructions in a suitable digital format or another accessible form. These instructions shall be concise, complete, accurate, and clear, ensuring they are relevant and understandable to the deployers (para 2.).

  \item \textbf{Article 14 Human Oversight} High-risk AI systems shall be designed with appropriate human-machine interface tools to ensure they can be effectively monitored by humans throughout their use (para. 1). Human oversight shall aim to prevent or minimize risks to health, safety, or fundamental rights that may arise from the risk within the system’s intended use or foreseeable misuse, primarily when other safeguards may not fully address these risks (para. 2). The human-machine interface shall also allow users to manually shut down the system in case of an emergency.

  \item \textbf{Article 15 Accuracy, robustness, and cybersecurity} High-risk AI systems shall be designed and developed to maintain appropriate levels of accuracy, robustness, and cybersecurity throughout their entire lifecycle (para. 1). To ensure these standards, the EU Commission, in collaboration with relevant stakeholders and organizations will promote the development of benchmarks and measurement methodologies for assessing accuracy, robustness, and other AI characteristic metrics (para. 2) \footnote{No guidelines or further clarification have been provided yet by the EU Commission.}.

\end{itemize}

\mypar{Article 17 Quality Management System} The QMS aims to structure and plan a process to control at least all listed regulations for high-risk AI systems. The QMS aims to ensure compliance with all AIA regulations, verify the AI systems design, and assure its quality. The QMS must be appropriately documented and contain at least the points listed in para. 1 lit. a - m.

\mypar{GPAI Models Regulations} A Summary of the key regulations for GPAI models \footnote{Chapter 5 Section 2 \& Section 3 AIA} is provided. For detailed provisions, refer to the complete articles in the AIA \cite{EU_AIA2024}. It is important to note that GPAI models can be included in high-risk AI systems. The obligations for GPAI models and high-risk AI systems must be fulfilled in such cases.

\begin{itemize}
  \item \textbf{Article 53 Obligations for providers of general-purpose AI models} Providers of GPAI models shall prepare and maintain up-to-date technical documentation of the model, including details of its training and testing processes and evaluation results  (para 1.). This documentation, which shall include at least the information listed in Annex \rom{6}, should be available to the AI Office and national competent authorities upon request (lit. a). Prepare, update, and provide documentation to other AI system providers who intend to integrate the GPAI model into their systems. This documentation shall be made available while respecting intellectual property rights and trade secrets under Union and national laws (lit. b).
  
  \item \textbf{Article 55 Obligations for providers of general-purpose AI models with systemic risk} In addition to the obligations outlined in Articles 53 and 54, providers of GPAI models with systemic risk shall evaluate the model using standardized protocols and tools that reflect current best practices, including adversarial testing to identify and mitigate systemic risks (para. 1 lit. a). Assess and address potential systemic risks, including their sources, arising from these AI models' development, market placement, or use (lit. b). Track, document, and promptly report severe incidents and corrective actions to the AI Office (lit. c). Ensure cybersecurity protection for the GPAI model and the system in which it is integrated (lit. d).
\end{itemize}

\mypar{Comments and Reviews} The AIA is legally binding for companies offering AI systems on the European market. However, according to some German legal experts, many uncertainties remain. According to \cite{bormhard2024_ai_act}, the EU AI Regulation needs more specificity in several areas, making concrete implementation unclear for many requirements. In \cite{schallbruch2021_ai_act}, the author highlights the high and numerous bureaucratic demands for high-risk AI systems. Due to this, compliance checks can require significant human and financial resources \cite{bormhard2024_ai_act}. Without innovative approaches to drafting these regulations, only large tech companies will likely have the resources to develop high-risk and generative AI systems. This situation could force small and medium-sized enterprises to leave Europe or avoid developing high-risk AI systems, contradicting the EU AIA Regulation’s objectives and stifling innovation.
Additionally, determining if an AI system is classified as high-risk can be very challenging, as described by \cite{ebers2024_ai_act}, for example, in the legal tech domain. Another concern is the AIA’s lack of alignment and consistency with other legal acts. In \cite{jaeckel2024_ai_act}, it points out that the EU AIA is not fully coordinated with, for example, the Medical Device Regulation, which governs medical devices, including medical software.

\subsection{Related Work}
\label{sect:pre:rw}
Literature was searched, published in legal-tech (such as \emph{JURIX}, \emph{AI \& Law}, \emph{ICAIL}, and \emph{AI and Ethics}), and information systems (such as \emph{ICPM}, and \emph{BPM}) journals and proceedings to identify theoretical, technical, and practical approaches to efficiently implement the AIA. 
Approaches for efficient, simplified risk class categorization under the AIA have been presented \cite{Hanif2023_though_decisions,golpayegani2023_high_risk}.
Some literature was found describing structured approaches to efficiently implement requirements for high-risk AI systems under the AIA. This includes strategies for Art. 9 AIA \cite{novelli2023_ai_risks,xia2023_towards_conc_ai_risk,tjoa2022_airman,nagbol2021_ris_assess_ai}, as well as methods to enhance AI explainability to make black-box models more transparent, supporting compliance with Art. 13 AIA \cite{sovrano2021_survey}  and Art. 14 AIA \cite{gorski2023_transparency_art14}. Additionally, work discussing technical metrics to quantitatively assess AI systems' risks and limitations are presented as required in Art. 9 and 15 AIA \cite{giudici2024_ai_risk_measure,bhaumik2023_audit,steimers2021_sources_of_risk}. Furthermore, research has been conducted to develop theoretical frameworks and processes for AI auditing, ensuring alignment with legal regulations like the AIA and ethical principles \cite{clarke2019_bpm_for_resp_ai,mökander2023_auditing_llm,ortega2023_ai_product_lifecycle,ellul2021_regulating_ai_regulator_perspective,simbeck2024_ai_auditing_fair_transparent}. The business process management community has taken up the  AIA in recent research. \cite{novelli2023_bpm_aia} use process modeling and execution to support reliable AI Fundamental Right Impact Assessment (cf. Art. 29a AIA). \cite{Pery2021_processmining_aia} emphasize the advantages of process mining for optimizing AI compliance management processes. Furthermore, practical AI compliance management systems like the proposed QMS approach have been introduced. For example, \cite{thelisson2024_careAI} present careAI, a web-based application that provides questions and checklists to evaluate an AI system’s risks, limitations, and benefits. \cite{floridi2022_capai} offer capAI, a framework to reliably structure and conduct AI system conformity assessments. In contrast to the design concept proposed in this paper, these tools are more akin to project management frameworks and tools and do not enable technical assessments.

%
%
%
%
\section{Quality Management System Design}
\label{sec:method}
The proposed QMS aims to support the AI compliance management process. This section explains the design and implementation in detail. First, the high-level requirements for structuring and implementing the QMS are outlined (cf. Sect. \ref{sec:meth/sub:req}). Second, the technical evaluation metrics incorporated into the prototype QMS to assess the connected LLM are defined. These metrics include three performance metrics, a gradient-based robustness metric, and a gradient-based explainability metric (cf. Sect. \ref{sec:meth/sub:tech_metr}). Next, the architecture of the QMS, aiming to be modular and scalable, is described, allowing integration with AI systems and incorporating several sub-services (cf. Sect. \ref{sec:meth/sub:arch}). The initial prototype QMS connects to the Phi-3-mini-128k-instruct LLM \cite{abdin2024_phi3} as an example of an AI system. It includes a risk management sub-service (Art. 9 AIA) and a data management and governance sub-service (Art. 10 AIA). LLMs were chosen for this prototype due to their popularity and the availability of pre-trained, open-source models on Hugging Face\footnote{Hugging Face: \url{https://huggingface.co}, accessed on 17 October 2024}. Lastly, the technical report presents the implementation results and the user interface of the QMS (cf. Sect. \ref{sec:meth/sub:proto}). The prototype's code can be found in the repository \url{https://github.com/henryk-mustroph/first_version_qmsAIA.git}, published on GitHub.

%
%
\subsection{Requirements}
\label{sec:meth/sub:req}
The QMS comprises functional (FR) and non-functional (NFR) requirements grouped into three distinct types. As defined in \cite{brugge2004_oo_se} (pp. 101-102), FRs describe the system’s functionalities, such as evaluating the AI system’s performance. In contrast, NFRs describe how the system should perform the functionality, such as ensuring that the AI system’s performance evaluation is done in under 2 seconds. The QMS incorporates \emph{Legal Requirements} (LR) derived from interpreting legal regulations set out in the AIA for high-risk AI and GPAI systems. Furthermore, \emph{System Design Requirements} (SDR) specify technical aspects of the QMS. These requirements cover \emph{human involvement}, \emph{architecture}, and \emph{computational and memory} needs. All requirements are high-level and will be subdivided into more concrete sub-requirements in future iterations. However, they represent the minimum necessary functionality for such a QMS and illustrate the extensive range of features that need to be integrated.

\mypar{Legal Requirements} Several requirements are derived from the given legal regulations. Articles 53 and 55 AIA provide information on requirements specifically for GPAI models, shown in Table \ref{tab:LR01}.

\begin{tinylongtable}{|p{1.0\linewidth}|}
\caption{Legal Requirements (GPAI Systems)} \\
\hline
\textbf{Legal Requirements} \\
\hline
\endfirsthead
\multicolumn{1}{c}%
{{}} \\ 
\hline
\endhead
\hline 
\multicolumn{1}{c}{{}} \\
\endfoot
\hline
\endlastfoot
\textbf{LR01: Article 53 Obligations for GPAI (FR):}\\
Draw up and maintain up-to-date technical documentation and provide transparency regarding the data used for the training, validation, and testing of the GPAI model
\\
\hline
\textbf{LR02: Article 55 Obligations for GPAI with systematic risk (FR):}\\
Implement evaluation metrics, using standardized protocols, incorporating robustness and security checks to assess and mitigate systematic risk.
\label{tab:LR01}
\end{tinylongtable}

The GPAI-specific requirements align closely with the requirements constructed based on the regulations for high-risk AI systems. The idea is to integrate all requirements for high-risk AI systems into the QMS. According to the EU AIA, the QMS should generally encompass strategies for regulatory compliance, including conformity assessment and management for modifications of legal guidelines or technical features. It should also provide techniques or procedures for the model's design, design control, verification, quality control, and quality assurance. Specifically, the QMS should contain an RMS to identify, analyze, assess, and mitigate potential risks and a DMDGS to demonstrate the data quality used for training, validation, and testing. The results should be stored in technical documentation. The software requirements derived from the obligation of high-risk AI systems are listed in Table \ref{tab:LR02} and must be ensured within the QMS.

\begin{tinylongtable}{|p{1.0\linewidth}|}
\caption{Legal Requirements (High-Risk AI Systems)} \\
\hline
\textbf{Legal Requirements} \\
\hline
\endfirsthead
\multicolumn{1}{c}%
{{}} \\ 
\hline
\endhead
\hline 
\multicolumn{1}{c}{{}} \\
\endfoot
\hline
\endlastfoot
\textbf{LR03: Article 9 Risk Management System (FR):}\\
Incorporate into the QMS a module for risk identification, analysis, and assessment functionalities, ensuring comprehensive coverage throughout the entire lifecycle of the AI system.
\\
\hline
\textbf{LR04: Article 10 Data and Data Governance (FR):}\\
Incorporate into the QMS a module to provide evidence that the data used for training, validation, and testing is unbiased, non-discriminatory, and compliant with privacy and data ownership regulations.
\\
\hline
\textbf{LR05: Article 12 Record keeping (FR):}\\
Develop an integration within the QMS to link and log the usage of the AI system in use, recording details such as the timestamp of usage, user identification, purpose of use, and the specific task for which the AI system is employed.
\\
\hline
\textbf{LR06: Article 13 Transparency Provision (FR):}\\
Incorporate into the QMS a transparency metric to assist users in interpreting the output of the AI system. This metric should provide clear and understandable insights into the decision-making process and underlying factors influencing the output.
\\
\hline
\textbf{LR07: Article 14 Human Oversight (FR):} \\
The QMS shall incorporate a UI for deployers and end-users of the AI system to document in-use risks and misuses and to have the control to shut up the system in emergencies.
\\
\hline
\textbf{LR08: Article 15 Accuracy, Robustness and Cybersecurity (FR):}\\
Implement into the QMS metrics to measure the system's accuracy, robustness, and cybersecurity.
\\
\hline
\textbf{LR9: Article 11 Technical Documentation (FR):}\\
The QMS shall allow to create and maintain up-to-date technical documentation for the AI system before its market release and after the market release.
\\
\hline
\textbf{LR10: Article 61 Post-Market Monitoring (FR):}\\
The QMS should be used to continuously assess the AI system's compliance with the requirements outlined in Chapter 2 after market release.
\label{tab:LR02}
\end{tinylongtable}

\mypar{System Design Requirements} The SDRs are categorized into three distinct types. First, The User Interface (UI) design and interaction specifications should encourage human involvement in checking and documenting AI systems. Second, architectural requirements should ensure a modular design, a smooth flow of data and communication between the services, and easy maintenance. The design should apply to multiple AI systems across various domains and tasks. Third, computational requirements should guarantee the efficient execution of technical evaluation metrics, even for large AI systems such as LLMs, to maintain optimal performance. According to \cite{ortega2023_ai_product_lifecycle}, the risk assessment process should adopt a human-centered design approach.
Moreover, besides human involvement in the RMS, users should be able to include references and descriptions of the data utilized for training, validation, and testing, consolidating all necessary information required by legal standards within the DMDGS. Additionally, users should be able to directly access and download technical documentation from the QMS to verify the test and validation proceedings on the AI system and ensure compliance with AIA regulations. Integrating all these functionalities into several sub-modules within a single tool aims to reduce effort, time, and costs for AI system providers and deployers. Table \ref{tab:SDR01} lists all UI and human involvement requirements.

\begin{tinylongtable}{|p{1.0\linewidth}|}
\caption{System Design Requirements (Human Involvement)} \\
\hline
\textbf{System Design Requirements} \\
\hline
\endfirsthead
\multicolumn{1}{c}%
{{}} \\ 
\hline
\endhead
\hline 
\multicolumn{1}{c}{{}} \\
\endfoot
\hline
\endlastfoot
\textbf{SDR01: User Interface (FR):}\\
The QMS shall provide a UI that actively engages users in the verification process of the AI system.
\\
\hline
\textbf{SDR02: Human-based Risk Management (FR):}\\
Pages to empower users to participate in the risk identification, analysis, assessment, and mitigation processes shall be included.
\\
\hline
\textbf{SDR03: Data Page (FR):}\\
Pages shall be included to upload or reference the data utilized for training, validation, and testing, accompanied by evidence of compliance.
\\
\hline
\textbf{SDR04: Downloadable Technical Documentation (FR):}\\
The QMS shall allow users to view and download the technical documentation for the AI system assessment.
\label{tab:SDR01}
\end{tinylongtable}

The QMS should have a modular design to react quickly to future changes in the legislative framework. The modular design is ensured by designing independent sub-services for each article in the AIA, which can be plugged into the QMS and maintained and updated separately at any time.
Key modules include the RMS (cf. Art. 9 AIA) and the DMDGS (cf. Art. 10 AIA), which are integrated into the first version of the prototype QMS. Potential additional modules, such as the AI system’s event logging (cf. Art. 12 AIA), can be added in future work.
The modules should be designed to apply generically to any type and architecture of AI system, except for the specific technical evaluation metrics, which need to align with the type and architecture of the underlying AI system. Additionally, the QMS allows users to persistently store technical documentation, past risk assessment processes, and data check references. This feature ensures comprehensive record-keeping and facilitates easy access to historical data, which is crucial for maintaining compliance and continuous improvement. The goal is to enable the QMS to be utilized throughout the entire lifecycle of the AI system, including post-market. All requirements of the system design architecture are detailed in Table \ref {tab:SDR02}.

\begin{tinylongtable}{|p{1.0\linewidth}|}
\caption{System Design Requirements (Architecture)} \\
\hline
\textbf{System Design Requirements} \\
\hline
\endfirsthead
\multicolumn{1}{c}%
{{}} \\ 
\hline
\endhead
\hline 
\multicolumn{1}{c}{{}} \\
\endfoot
\hline
\endlastfoot
\textbf{SDR05: Modular Design (NFR):}\\
Enable users to customize and modify the QMS modules, add various AI systems, and incorporate technical metrics tailored to domains, purposes, and tasks.
\\
\hline
\textbf{SDR06: Communication and Data Flow (NFR):}\\
Implement reliable and secure data communication across the various modules of the QMS.
\\
\hline
\textbf{SDR07: Persistent Storage (NFR):}\\
Set up a database to persistently store all user information, technical documentation, and assessment processes.
\label{tab:SDR02}
\end{tinylongtable}

Identifying the required computational resources is the third type of system design requirement. LLMs, for example, require significant computational power for tasks like calculating the gradients needed for some technical evaluation metrics. These gradients are necessary for certain adversarial attacking and explainability techniques. The specific GPU and CPU resources must be tested, evaluated, and defined for a second version of the prototype QMS. Table \ref{tab:SDR03} details the computational requirements for the system design.

\begin{tinylongtable}{|p{1.0\linewidth}|}
\caption{System Design Requirements (Computation)} \\
\hline
\textbf{System Design Requirements} \\
\hline
\endfirsthead
\multicolumn{1}{c}%
{{}} \\ 
\hline
\endhead
\multicolumn{1}{c}{{}} \\
\endfoot
\hline
\endlastfoot
\textbf{SDR08: Computational Resource (NFR):}\\
Allocate sufficient computational resources, including GPU resources, to ensure that cost-intensive computations, even on large GPAI models, can be computed within an acceptable timeframe.
\\
\hline
\textbf{SDR09: Performance (NFR):}\\
Implement high-performance technical evaluation metrics for cost-intensive computations that minimize GPU storage usage and improve execution time.
\label{tab:SDR03}
\end{tinylongtable}

%
%
\subsection{Technical Evaluation Metrics}
\label{sec:meth/sub:tech_metr}
Conducting technical metrics on the AI system is essential to comply with the AIA regulations, such as shown by \cite{giudici2024_ai_risk_measure} for Art. 9 AIA to assess risks, or \cite{gorski2023_transparency_art14} for Art. 14 AIA  testing the AI system's transparency and mandatory in Art. 15 AIA to assess the AI system's accuracy and robustness. Since the first version of the prototype connects to an LLM, technical metrics for neural networks, specifically LLMs, are defined and implemented. Several metrics already exist to assess classification or regression models on performance, explainability, robustness, or fairness, as presented in works by \cite{giudici2024_ai_risk_measure,nagbol2021_ris_assess_ai,bhaumik2023_audit}. However, evaluating generative models on these characteristics presents unique challenges. The technical metrics presented are specifically designed to evaluate the \emph{performance}, \emph{explainability}, and \emph{robustness} of the connected Phi-3-mini-128k-instruct on specialized tasks. The proposed specialized metrics are optimized for text summarization and transformation tasks. This selection aligns with the application of LLMs in automated compliance verification approaches in previous works (cf. \cite{Barrientos2023_temp_comp,henryk2023_resource_comp,henryk2023_snm_comp}), which could have a strong use case in potential high-risk domains such as the legal and financial domains. This selection has been chosen as an example based on previous research. Any other tasks and technical metrics could have been chosen.

\mypar{Basic Definitions} A text has always a domain and a type, for example, a legal text in the medical domain. The text \(t\) consists of \(n\in\mathbb{N}\) tokens: \( t = t_0, \ldots, t_{n-1}\). The LLM calculates with numerical values. For that, each token is encoded and embedded in a numerical vector which is the input for LLM inferences. Hence, the encoding function and decoding functions are defined as: 
\begin{align*}
enc\colon& \text{token} \to \mathbb{R}^{\text{embed dim}}, & dec\colon&  \mathbb{R}^{\text{embed dim}} \to \text{token}.
\end{align*} 
The encoded tokens are then defined as: \( enc(t) = x\). 
An LLM is defined as a function \(f\) that takes a sequence of \(n\) embedded tokens and returns a sequence of \(m\) embedded tokens:
\[f\colon \mathbb{R}^{n \times \text{embed dim}} \to \mathbb{R}^{m \times \text{embed dim}}, \; x \mapsto \Tilde{y}\]
The logit values \(\Tilde{y}\) outputted by the LLM can be decoded to obtain the generated output tokens: \(dec(\Tilde{y}) = \Hat{y}\).
Additionally, each input text has a predefined ground-truth target output: \({y}\).

\mypar{Performance} According to \cite{banerjee2023_benchmark_llm}, one generic and always efficient performance metric is calculating the LLM's accuracy score for a specialized task. Since we concentrate on the evaluation of summarization and text transformation, true positives (TP) are defined as a set of actor-activity pairs that are present in the predicted output text \(\hat{y}\) and the ground truth text \(y\). True negatives are defined as a set of actor-activity pairs that are not present in \(\hat{y}\) and not in \(y\). False positives are defined as a set of actor-activity pairs that are not present in \(\hat{y}\) but in \(y\). False negatives are defined as a set of actor-activity pairs that are present in \(\hat{y}\) but not in \(y\). Furthermore, as noted in \cite{banerjee2023_benchmark_llm}, the BLEU score (described in \cite{Papineni02_bleu}) is typically employed for assessing language model performance in translation tasks. In contrast, as detailed in \cite{lin2004_rouge}, the ROUGE score is the more suitable metric for language model performance assessment in summarization tasks. Several ROUGE methods exist. The chosen one here is the ROUGE-N method.
\[\operatorname{ROUGE-N} =  \frac{|\operatorname{n-gram}(y) \cap \operatorname{n-gram}(\Hat{y})|} {|\operatorname{n-gram}(y)|}\]
A n-gram is defined as: \(\operatorname{n-gram}(t_0, ..., t_{m-1}) = \{t_i, t_{i+1}, ..., t_{i+n-1} \mid  0 \leq i \leq m-n\}\). Here, we defined the recall ROUGE-N, but the precision pendant and the F1 ROUGE-N score are also implemented. The following gives an example of how the n-grams are constructed. \\
\(t =\) \textit{The user creates a new process instance, then the system can execute the instance and stop it afterward.}
\(\Hat{y} = \)  [``User'', ``creates new process instance'', ``system'', ``execute instance'', ``system'', ``stops instance'']. 
\(\operatorname{2-gram}(\Hat{y}) = \) [``User creates new process instance'', ``creates new process instance system'', ``system execute instance'', ``execute instance system'', ``system stops instance''].
Another implemented performance metric, explained in \cite{Hu_2020_sys_asses_neural_lm}, which is a general performance metric for neural network-based architectures, is the Perplexity score. It computes the uncertainty of the LLM for predicting each subsequent token in the output sequence. A lower perplexity value signifies less uncertainty and, thus, a higher confidence in the correctness of the prediction.
\[\text{PP}(f, t) = \left( \prod_{i=0}^{n-1} \frac{1}{P(\Tilde{y}_i | \Tilde{y}_{i-1}, \dots, \Tilde{y}_{0})} \right)^{\frac{1}{n}}\]
Thereby, the \(P(\Tilde{y}_i | \Tilde{y}_{i-1}, \dots, \Tilde{y}_{0})\), is the conditional probability of each next token outputted by the LLM.

\mypar{Explainability} The explainability of black-box models such as neural networks and, more specifically, huge neural networks such as LLM is still a hot topic in ongoing research. However, three distinct general explainability approaches exist for LLM. Perturbation-based, gradient-based, and self-explanation-based methods. Examples of perturbation-based methods are occlusion and LIME \cite{huang_2023_self_explain_LLM,krishna2023_post_hoc}. Occlusion measures the difference between a generated LLM output based on the entire input sequence and the output based on the same input sequence with one word deleted \cite{li2016_occlusion}. LIME removes a random subset of words from the input sequence, then computes the model's output and compares the output with the model's outputs by removing just one word from the output. They compute a linear regression model comparing the relationship between the different input and output types \cite{tulio2016_lime}. Secondly, gradient-based explainability methods exist. One method originally invented for image classifiers by \cite{simonyan2014_saliency} is the generation of saliency maps. Originally, saliency maps aimed to visualize and rank pixels of an input image based on their sensitivity and influence on the predicted output class. \cite{krishna2023_post_hoc} also employed the gradient-based approach. They devised a post-hoc explanation technique by formulating a meta-prompt using the most significant input tokens to produce the specified output. This method aims to enhance prompting language models by comprehending the model's reasoning behind generating a particular output. It involves computing the gradients of the LLM's target output and each input token and selecting the top-k tokens that exerted the most influence on the model output. The third approach involves LLM self-explanation methods. \cite{huang_2023_self_explain_LLM} delineate two distinct types of LLM self-explanation. Firstly, the explanation-prediction method requires the model to identify the most pertinent tokens in the input sequence and subsequently generate a prediction based on them. Secondly, the prediction-explanation method necessitates the model to provide a prediction and then elucidate the rationale behind that prediction. This becomes interesting for models like GPT-4, which are not openly available, making it impractical to compute perturbations and gradients. 

In this prototype, a \textit{gradient-based saliency map metric} is implemented since gradient-based methods are the most promising in getting the hard facts on how the model processes the input.
\[S = { \frac{\delta \Tilde{y}}{\delta x}} = \{\sigma_0, \dots, \sigma_{n-1}\}\]
Here, \(S\) denotes the set of gradient values for each input token based on the model output. In a second step, all computed gradient values are normalized: \( \lVert S \rVert = \{\lVert \sigma_0 \rVert, \dots, \lVert \sigma_{n-1} \rVert\}\). Each normalized gradient value ranges from 0 to 1, with values closer to 1 exerting a stronger sensitivity and therefore influence on the predicted output than those at 0. Utilizing these normalized gradient values, the corresponding input tokens are visualized. Tokens with greater influence are depicted in lighter shades of green, while those with lesser influence appear in darker shades of red. This approach produces a token-based saliency map, providing users insight into which input tokens have the most significant impact on the generated output.

\mypar{Consistency} In this prototype, a \textit{gradient-based adversarial examples metric} is chosen to be implemented as a consistency method. The overall idea is to measure the degree and effort of changes in the input text required to generate a predetermined false output. This concept is related to generating gradient-based adversarial examples (cf. \cite{szegedy2014_intriguing,goodfellow2015_adversarial_att}) and is a robustness and security check for neural networks. The consistency metric is based on a gradient-based adversarial example creation for LLM, similar to the concept introduced by \cite{yao2023llm_lies}. This approach utilizes adversarial example creation to induce hallucinations in the LLM's output, effectively evaluating its robustness. Mathematically, it is similar to the gradient-based explainability metric. It involves modifying the text input data by modifying the top-k most important tokens until the LLM generates with high certainty a similar or even the predetermined false output containing hallucinations. The robustness, reliability, and security of the LLM can be evaluated by analyzing the type and quantity of replaced words and the number of replacement iterations. First, the gradient values \(S\) are calculated for each token as in the presented gradient-based explainability metric. In the second step, the input tokens are modified by a factor \(\eta\), which can be set and adjusted, similar to the learning rate in neural network optimization. The factor must be added in the direction of the gradient value. The goal is to modify the input in a direction to fool the LLM such that the lowest loss for \(f(x)\) 
\[x = \sum_{i=0}^{n-1} x_i + \eta \cdot \operatorname{sign}(\sigma_i)\]
The embedded input token modification is done once the LLM outputs the predetermined adversarial target. The model's input from the previous iteration is modified in each iteration to calculate the gradients.

Gradient-based metrics were chosen for explainability and consistency (robustness) due to several important factors. These metrics are fundamental for understanding a model’s output and sensitivity to input data. They offer crucial insights into the model’s architecture and reasoning. Additionally, gradient-based metrics are among the most challenging to implement and compute, especially considering the required GPU memory. This aspect was particularly relevant for the prototype, as it allowed for the analysis of the system’s computational and memory needs.

%
%
\subsection{Architecture and Data Models}
\label{sec:meth/sub:arch}
\begin{figure}[!ht]
    \centering
    \includegraphics[width=1.0\textwidth]{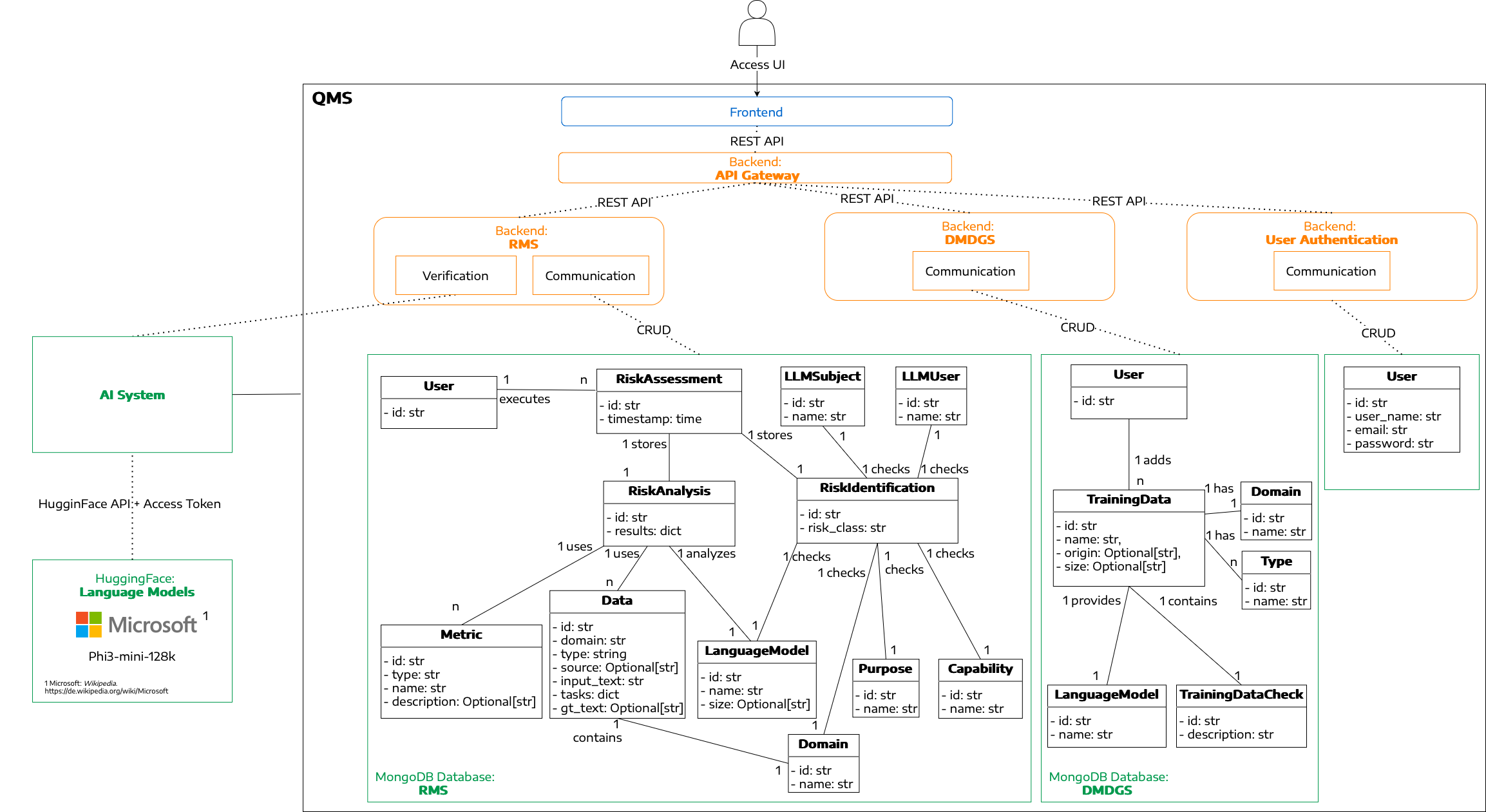}
    \caption{Architecture and Data Models of the QMS}
    \label{fig:architecture}
\end{figure}

The QMS architecture is based on a microservice design (cf. \cite{nadareishvili2016_microservice}). The current prototype QMS consists of an RMS module, a dedicated DMDGS module, and a user authentication module, which only stores user and login information. All modules are independent sub-services containing their backend and database. The architecture of the QMS and the data models for the RMS, DMDGS, and user authentication databases as UML class diagrams are illustrated in Fig. \ref{fig:architecture}.

Users can interact with the UI implemented on the frontend to participate in the verification and documentation process within the QMS by executing risk assessments and ensuring data management and governance compliance. An API Gateway backend service orchestrates all user requests from the frontend and forwards each request to the corresponding backend system. The API Gateway loads an environment file containing the names or pseudonyms for each sub-service root. The backend of the corresponding sub-service then sends the response back to the API Gateway, which forwards the message to the frontend, where it is loaded into the UI for the user. This structure, data transfer, and orchestration reduce complexity and improve modularity. Complexity is reduced, and modularity is ensured because adding sub-services to the QMS does not alter existing data transfers. To add a new service to the communication and data transfer pipeline, only an additional entry in the environment file read by the API Gateway is required. This prototype QMS focuses on the functionalities of the RMS rather than those in the DMDGS, as already mentioned. The RMS backend consists of two different components. It contains a verification component that can load different language models available from Hugging Face and implements all technical evaluation metrics presented in the next section (cf. Sect. \ref{sec:meth/sub:proto}), which are used to analyze and quantitatively assess the LLM's performance, consistency, and explainability for specialized tasks in selected high-risk domains. Furthermore, the RMS backend contains a communication component responsible for frontend-backend-database communication and data transfer. Each sub-service, including the DMDGS and user authentication service, contains a communication component with the same structure and functionality. In contrast to the RMS, the DMDGS, and the user authentication services only include a communication component and no further components. The RMS data model consists of three main classes: risk identification, analysis, and assessment. A risk assessment object stores one risk identification and one risk analysis object. Users can execute multiple risk assessments, with the risk assessment IDs stored in the user object. This setup ensures that users can quickly review past risk assessment results in the UI. The DMDGS data model follows the same principle. The user class stores a list of data IDs, where each data object is linked to an LLM object and a data check object that verifies the compliance of the added data reference object. It is important to note that the RMS risk analysis requires data, which has a domain and task to assess the LLM's characteristics, efficiency, and risks (performance, explainability, and consistency). On the other hand, the data referenced in the DMDGS are the training, validation, and testing data to train and develop the LLM, which will be assessed in the RMS after it is ready to use. The user authentication data model only contains a user class to store user IDs, personal data, and login information. The following technology stack is used to implement the prototype QMS: The frontend is developed using JavaScript and the React.js \footnote{React.js: \url{https://react.dev}, accessed on 17 October 2024.} library. All backend services are written in Python, utilizing various libraries and packages such as PyTorch \footnote{PyTorch: \url{https://pytorch.org}, accessed on 17 October 2024}, Transformers \footnote{Transf.: \url{https://huggingface.co/docs/transformers/}, accessed on 17 October 2024} in the verification component to load the LLMs from Huggingface into the GPU and to perform computations on the loaded LLMs, and FastAPI \footnote{FastAPI: \url{https://fastapi.tiangolo.com}, accessed on 17 October 2024} and  PyMongo \footnote{PyMongo: \url{https://pymongo.readthedocs.io/en/}, accessed on 17 October 2024} to implement all REST and CRUD methods to provide a communication pipeline from the frontend to the backend and from the backend to the database. MongoDB \footnote{MongoDB: \url{https://www.mongodb.com}, accessed on 17 October 2024}, a non-SQL database, is employed for data storage, allowing for rapid and straightforward modifications and design changes. MongoDB uses so-called collections (database tables), which can be modeled and designed using UML class diagrams and can always be extended or changed, keeping them flexible. The documents are the elements of a collection (rows of a table). They are stored in the collections as JSON data, maintaining a unified structure and language used in the frontend. A document can contain different collections of document data, similar to object-oriented programming, making MongoDB easy to understand.

\begin{figure}[!ht]
    \centering
    \includegraphics[width=1.0\textwidth]{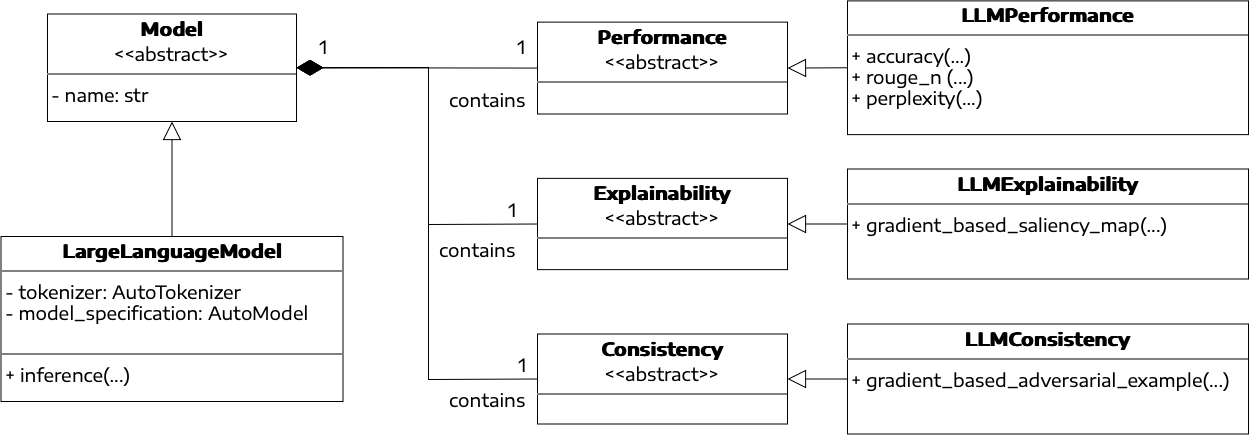}
    \caption{Architecture of the RMS Verification Component}
    \label{fig:verif_comp}
\end{figure}

\mypar{RMS Verification Component} A special design approach is chosen to implement the technical evaluation metrics in the verification component of the RMS. 
The QMS should be adaptable to every type of AI system, ensuring that technical tests can be conducted on each AI system. The Strategy Pattern \cite{brugge2004_oo_se} (p. 506), a behavioral software design pattern, has been used to ensure a flexible design. Fig. \ref{fig:verif_comp} depicts the implementation design of the RMS verification component. 

The Strategy Pattern contains a context and a strategy. The strategy is abstract and contains several children with concrete strategy implementations. Then, the context can choose the best-fitting strategy for its task at runtime. The applied Strategy Pattern is modified with minor changes in the RMS verification component. The context is also an abstract class; concrete context child classes implement the context logic. The \emph{Model} class is the abstract context class and represents the AI systems. Only the concrete context class \emph{LargeLanguageModel} exists, but additional, such as a Linear Regression Model class, can be easily extended. The abstract \emph{Model} class is connected to the abstract strategy classes: \emph{Performance}, \emph{Explainability}, and \emph{Consistency} via a composition, meaning that the abstract strategy can only exist as long as the model class exists. The \emph{Model} class contains a \emph{Performance}, \emph{Explainability}, and \emph{Consistency} object in its constructor, which are initialized and further defined in its child class \emph{LargeLanguageModel}. Specifically, the \textit{LargeLanguageModel} class contains \emph{LLMPerformance}, \emph{LLMExplainability}, and \emph{LLMConsistency} objects as concrete strategies in its constructor. The concrete strategy classes contain methods for each technical evaluation metric. Generally, this implementation design guarantees modularity and makes it easy to add new concrete contexts (types of AI systems), strategies (AI characteristics), and concrete strategies (technical evaluation metrics) without changing existing code.

%
%
\subsection{Prototype QMS - First Version}
\label{sec:meth/sub:proto}
The prototype QMS is accessible at \url{https://power.bpm.cit.tum.de/qmsAIA/}.

\begin{figure}[H]
    \centering
    \includegraphics[width=0.7\textwidth]{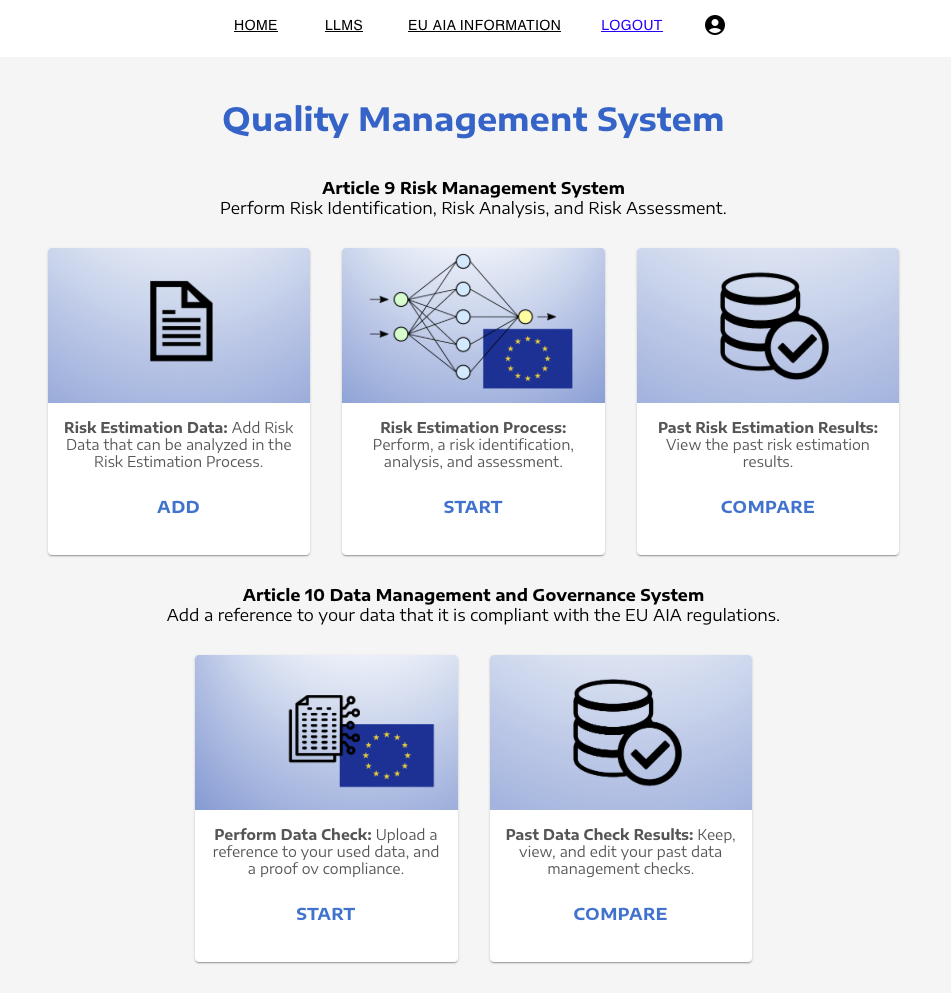}
    \caption{QMS - Home}
    \label{fig:qms:home}
\end{figure}

\mypar{Main Service: Quality Management System} The user accesses the QMS home page after signing up and signing in. The prototype QMS home page design is depicted in Fig. \ref{fig:qms:home}.

To sign in, the user must enter \emph{test} as username and password. The homepage is subdivided and structured into the prototype's several sub-services. Each functional sub-service (RMS and DMDGS) is listed in a separate row, one below the other, and contains several boxes each, one for performing risk assessments or data checks and one for viewing past risk assessments or data checks. Additionally, the UI provides access to various sections with details about the LLM and the AIA.

\mypar{Sub-Service 01: Risk Management System} As mentioned, the main focus of the prototype QMS was on the RMS sub-service. The overall structure of the RMS is based on the ISO 31000 standard \cite{ISO31000}, which is also applied to the AI system RMS presented in \cite{tjoa2022_airman}. It follows a \textit{``Plan-Do-Act-Check''} principle: Plan involves planning the RMS based on the specific use case and defining the processes' sub-activities, Do encompasses conducting the designed RMS process, Act entails inspecting and evaluating the entire RMS process, and output, and Check involves refining and optimizing the RMS process. Additionally, \cite{nagbol2021_ris_assess_ai} establishes five design principles for RMS for AI systems, which should also be integrated. These principles include i) incorporating multi-perspective expert assessment, such as drawing insights from various domains and involving AI experts in the process, ii) encouraging participation from diverse stakeholders in the risk assessment process, iii) identifying risks based on real-life scenarios, and iv) analyzing risks using metrics beyond accuracy, v) ensuring ``human in the loop'' processes for black-box models. This RMS can be seen as a prototype within the prototype QMS. The construction process of this RMS, according to ISO 31000, consists of an iterative process containing six sub-activities: component selection, risk identification, verification data selection, risk analysis, risk assessment, and risk mitigation.
In the component selection, the users should be encouraged to select the analyzed AI system (in this case, the LLM) and the task the AI system performs. In the second step, the user performs risk identification based on the risk categorization strategy for AI systems presented by \cite{golpayegani2023_high_risk}. A vocabulary-based approach is used where the user adds values for the LLM's domain, purpose, capabilities, LLM user, and LLM subject. Based on these selected values, the risk class of the AI system is determined. A basic algorithm was implemented to evaluate the risk class according to the categorization presented by \cite{golpayegani2023_high_risk}. Future work will explore changing the risk identification process to a more stakeholder-oriented approach, as mentioned in the design principle by \cite{nagbol2021_ris_assess_ai}.
Additionally, it will be considered whether risk identification should be designed to identify foreseeable risks and misuses, assuming the AI system is already classified as high-risk, instead of categorizing the AI system into a risk class. In the third step, the user can quantitatively analyze the AI system’s performance, explainability, and consistency risks. The user can choose between different technical evaluation metrics from the risk analysis to the selected LLM. In the fourth step, the model calculates a risk assessment documentation based on the risk identification and analysis results. This documentation aims to verify compliance with AIA regulations. The current documentation will not serve as proof. However, the basic principle and the feasibility of automatically creating and downloading a risk and technical documentation from the prototype QMS can be demonstrated. In the final step, users can add strategies to mitigate assessed risks. This feature must still be fully developed and integrated into the prototype's second version.

The process was demonstrated with test data to provide an example of how such a risk analysis and assessment result looks, and all presented metrics were selected to be computed. As LLM, the Phi-3-mini-128k-instruct from Microsoft containing 3.8 billion parameters was analyzed on the domain \emph{``Industry Process Description''}, and the task \emph{``Summarization''}. The model input was: \emph{``Extract the Actor and Activity pairs from the text. Return only the list of JSON documents in the following format: [{'actor': 'example\_actor\_1', 'activity': 'example\_activity\_1'}, {'actor': 'example\_actor\_2', 'activity': 'example\_activity\_2'}, ...] without any further explanation: The user creates a new process instance, then the system can execute the instance and stop it afterward.}''.

\begin{figure}[H]
    \centering
    \includegraphics[width=0.9\textwidth]{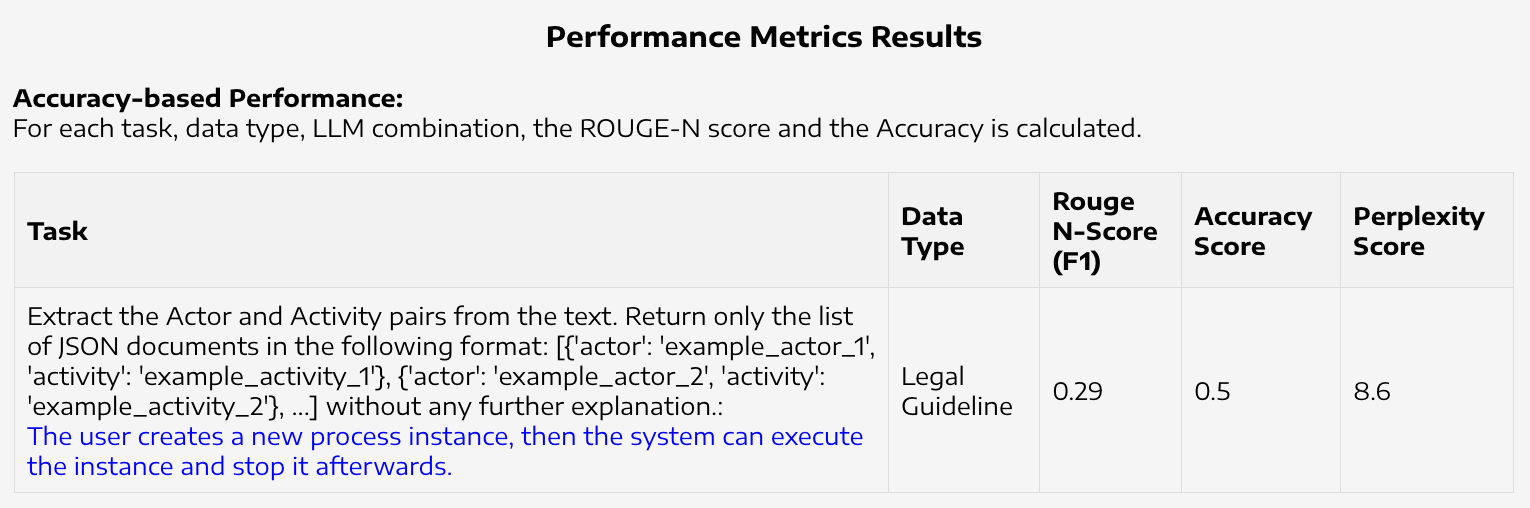}
    \caption{QMS - RMS - Assessment - Performance Result}
    \label{fig:app:risk_ass_acc}
\end{figure}

In Fig. \ref{fig:app:risk_ass_acc}, the performance results of the model from the risk assessment UI page are shown. The performance results display the numerical values for the model’s accuracy, Rouge score, and perplexity for the given input tasks.

\begin{figure}[H]
    \centering
    \includegraphics[width=0.9\textwidth]{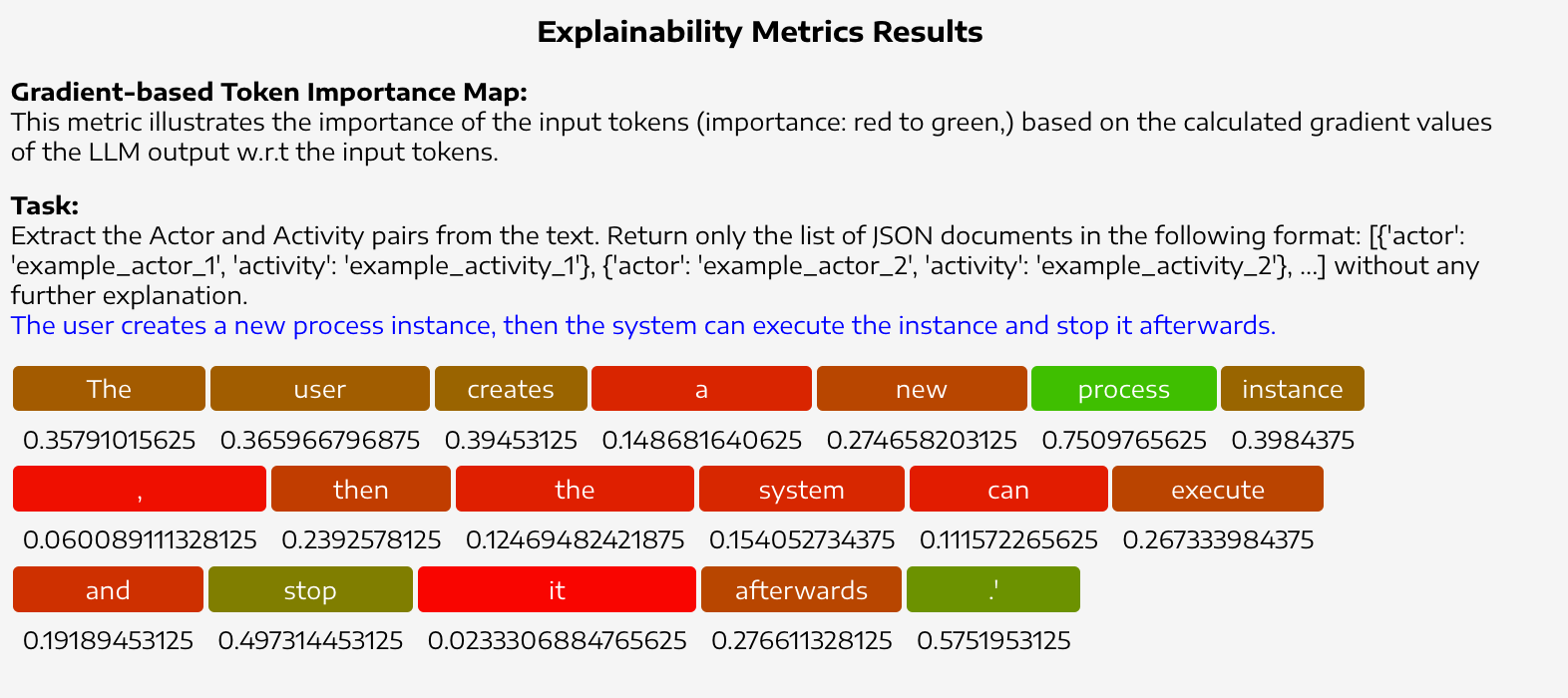}
    \caption{QMS - RMS - Assessment - Explainability Result}
    \label{fig:app:risk_ass_exp}
\end{figure}

In Fig. \ref{fig:app:risk_ass_exp}, the explainability result is visualized as a saliency map, where each input token is colored between red (unimportant, less sensitive) and green (important, sensitive) to the model’s generated output.

\begin{figure}[H]
    \centering
    \includegraphics[width=0.9\textwidth]{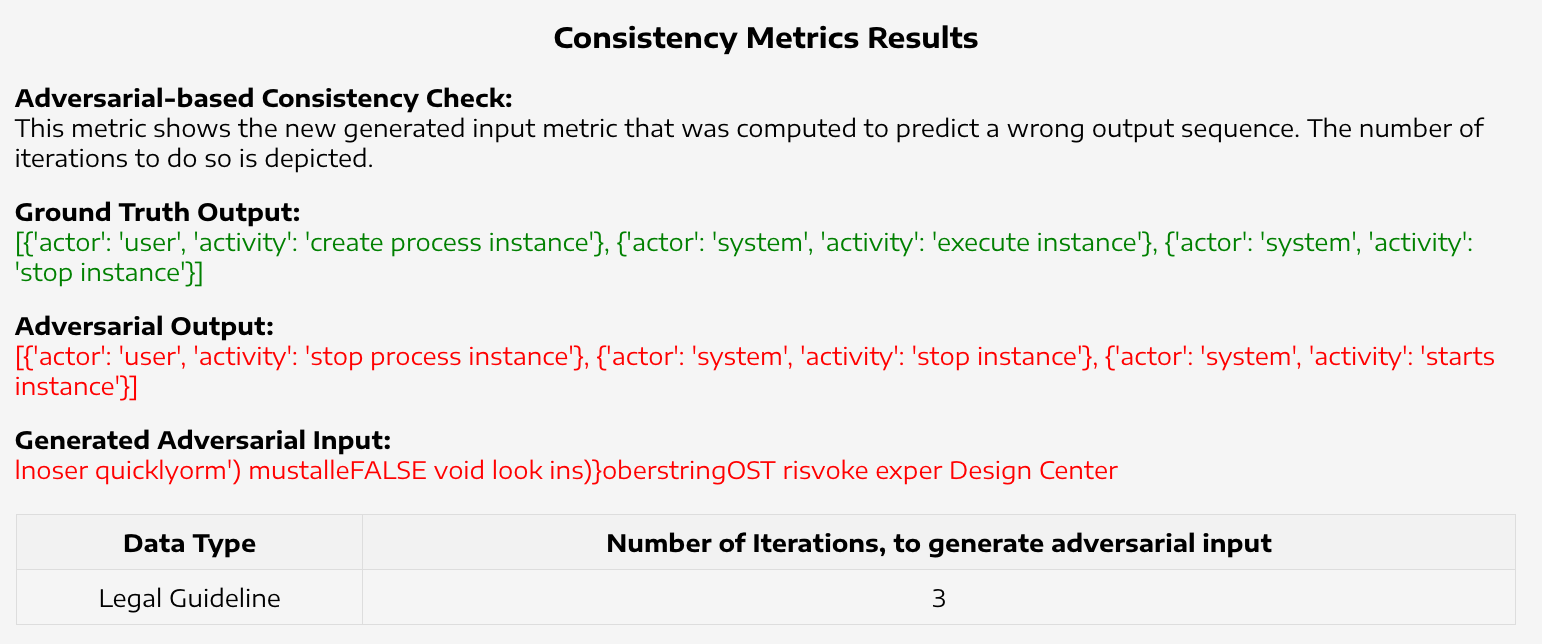}
    \caption{QMS - RMS - Assessment - Consistency Result}
    \label{fig:app:risk_ass_cons}
\end{figure}

The consistency result is depicted in Fig. \ref{fig:app:risk_ass_cons}. The consistency results display the ground-truth output (green), the adversarial output (red) generated by modifying the input tokens in the gradient direction using a hyperparameter similar to the learning rate, and the number of iterations to fool the model.

\mypar{Sub-Service 02: Data Management and Governance System} The DMDGS is the second sub-service in the prototype QMS and is kept basic. It consists of two components in the UI: the data check page and the past data check page. The former allows the user to select the AI system (in this case, LLM) for which training, validation, or testing data has been used. Furthermore, the user can specify the name of the dataset, the origin, the type, the domain, and the size. To prove that the data is checked and compliant under Art. 10 AIA, the user has a text field in which he/ she can enter a textual reference to the proof. In future work, the goal is to enhance the Data Management and Governance System (DMDGS) to provide functionality for analyzing and assessing datasets for training, validation, and quantitative testing within the QMS. This includes evaluating datasets for biases and errors. Additionally, the QMS should verify that dataset splits are well-determined and that the data aligns with the goals of AI system development. It is also essential to check whether other EU data regulations, such as the GDPR, must be considered within the QMS.

%
%
%
%
\section{Evaluation}
\label{sec:evaluation}
First, the requirements and implementation of the initial prototype are evaluated. Next, the design of the prototype is assessed through a workshop with IT and AI experts and an interview with a legal expert.

%
%
\subsection{Requirements}
\label{sec:evaluation:req}
The proof-of-concept prototype demonstrates the feasibility of the proposed QMS design concept that connects to an AI system for technical assessments and incorporates multiple sub-services, each serving a specific purpose. Human involvement is facilitated through a web-based, service-oriented application. The QMS prototype utilizes a microservice architecture, enabling a modular and loosely coupled design to integrate multiple sub-services. The first-version prototype QMS incorporates no standard protocol requirements derived from the ISO 4213 \cite{ISO4213} for AI performance evaluation, ISO CD TS 6254 \cite{ISO6254} for AI explainability evaluation, and ISO TR 24029 for AI robustness evaluation \cite{ISO24029}. However, these might be relevant to be compliant with more technical AIA requirements such as Art. 15 AIA. The metrics and calculations are currently included in the risk analysis component. However, a different sub-service can also be created for LR08 (Article 15 EU AIA) to prove the AI system's accuracy, robustness, and cybersecurity separately. Computational and memory demands pose challenges, particularly for huge AI models like LLMs. For instance, the Phi-3-mini-128k-instruct, with 3.8 billion parameters, requires \(2 \cdot 3.8 \cdot 10^{9} \: parameters \cdot 4 \: bytes = 30.4 \cdot 10^{9} \: bytes\) = 30.4 GB of memory for a single forward and backward pass at full float32 precision. Reducing precision to float16 halves this requirement. The NVIDIA RTX 4090 GPU, with 24 GB of memory, can accommodate this LLM, which requires 15.2 GB in float16 to compute the implemented robustness and explainability metrics. Future work will focus on developing comprehensive, complete, and verifiable sub-services. To achieve this, ontology-based, vectorized database schemes may be implemented as proposed by \cite{hernandez2024_mapping_standard_aia}, enabling the detailed design of sub-services tailored to AIA requirements and enriched with relevant information from associated ISO standards. This approach will facilitate the implementation and execution of verified processes through workflow engines, as advocated by \cite{novelli2023_bpm_aia}.

%
%
\subsection{Workshop with IT and AI Experts}
\label{sec:evaluation:workshop}
A workshop was conducted to evaluate the prototype following the guidelines and principles of \cite{thoring2020_workshop}. 

\mypar{Method} Seven experts in information systems, business process management, and AI participated in the workshop. Three of them are continually involved with AI, especially LLMs, in their research. First, a presentation and prototype QMS live demo were provided, continuing with a focus-group discussion. The goal of the workshop was to gather feedback on the first version of the prototype QMS and to identify additional requirements for the RMS or the related future sub-service that allows them to technically assess the AI systems such that the QMS can be used to verify, evaluate and document AI systems entirely on different tasks. Therefore, the focus was more on evaluating the integrated RMS, technical metrics, and assessment capabilities. 

\mypar{Result} Five of seven experts agreed that such a tool would be helpful and increase efficiency when technically evaluating AI systems on specialized tasks, even if they mentioned that complex tasks might require manual inspection. LLMs whose output cannot have a pre-defined ground truth could be incredibly challenging for the QMS. Two had no opinion on such a tool since they never researched or worked on AI and needed to be more familiar with the AIA. Additionally, helpful technical requirements are elicited to improve the prototype QMS's usefulness: i) the prototype QMS users shall be able to upload and store their data to analyze and assess the risk of the AI system (cf. Art. 9, 15 AIA). The data referred to by the experts pertains solely to the data used for evaluating the AI system during the post-development \& pre-release phase and is unrelated to the training data. ii) Comprehensive data buckets of similar data should be automatically constructed based on the type and domain of data QMS users add. iii) Complete coverage of test cases is essential, meaning that all possible scenarios and edge cases must be considered and tested to guarantee the reliability and effectiveness of the AI system. This implies that the data used for evaluation should be comprehensive, mostly error-free, and cover real-life settings. iv) Clarification shall be needed, noted, and provided to the QMS user regarding manual procedures that cannot be automated within the QMS.

%
%
\subsection{Inteview with Legal Expert}
\label{sec:evaluation:interview}
To evaluate the prototype, an interview was conducted with a legal expert and a corresponding qualitative content analysis (QCA) following \cite{glaser2009_interview}.  

\mypar{Method} The QCA consists of a structured process containing four main steps: First, a research question (pp. 61-70) and a theoretical preliminary idea are constructed containing hypotheses built on a search pattern (pp. 73-90). The search pattern contains independent (implementation and conformance with AIA regulations, technical design, and incorporation of other EU regulations), intermediate (Prototype QMS), and dependent (ease verification and documentation of AI, improve efficiency in AI Compliance Management) variables. Based on having causal mechanisms between independent, intermediate, and dependent variables, a hypothesis was constructed: \emph{The implementation of AIA regulations as software requirements and the architecture influence the design of the prototype QMS, which can increase the efficiency for companies in the AI compliance management process} which is either proved or disproved as a result of the QCA. Second, the interview partners are acquired, and an interview guideline is set up (pp. 111-150). Third, expert interviews are transliterated, extracted, analyzed, and categorized into the search pattern categories (pp. 197-257). Fourth, the results are interpreted and presented (pp. 261- 275). 

\mypar{Result} The legal expert evaluated the prototype positively. It was mentioned that the QMS has a very good, precise design and is, therefore, easy to use, even for auditors with legal expertise but possibly less AI expertise, such as potential auditors for national authorities or internal compliance team members and auditors. Therefore, one legal expert confirmed the hypothesis that the design concept and the prototype QMS positively impact AI systems' verification, documentation, and compliance management process. Furthermore, it was confirmed that the QMS should be broad enough to accommodate, audit, and document all types of AI systems, including those not classified as high-risk. This approach would simplify the internal compliance management process for any AI system. Additionally, helpful technical requirements are elicited to improve the prototype QMS usefulness: i) The QMS shall technically check and verify AI systems according to generalized standard protocols such as ISO standards. ii) The QMS should adapt to changes in AI systems by focusing checks solely on modifications, with the results integrated into the existing documentation. iii) Documentation (cf. Art. 11 AIA) and potentially the transparency obligations (cf. Art. 14 AIA) shall not be displayed as separate, independent sub-services. Instead, all other sub-service results shall be selectable and added to the technical documentation. The technical documentation shall be a collection of all sub-service results. iv) Alternating reactions between the AI system, the AIA, and other harmonization standards (cf. Annex I AIA) depending on the domain of the AI system could arise and need to be checked.  

To further enhance the prototype QMS, future work will expand qualitative evaluations by engaging a larger group of IT, AI, and legal experts. This will build on the insights provided by evaluating the first version of the prototype QMS, which only involves seven IT and AI experts and one legal expert. The opinion of further legal experts would help to refine the requirements and usability of the QMS and manifest the constructed hypothesis.

%
%
%
%
\section{Conclusion}
\label{sec:disc}
The presented QMS is a tool for ensuring compliance with the AIA regulations for high-risk AI and GPAI systems, guided by legal and system design requirements. The QMS is based on a microservice architecture, directly connecting AI systems and containing several sub-services, each with a different purpose based on the regulations of the AIA. This SaaS web application aims to map the compliance management processes for AI systems into one tool and to carry them out efficiently. Assessments can be directly applied to the adopted AI system within the QMS. Although the first version of the presented prototype QMS is optimized to check and document LLMs, its concept, design, and architecture can be applied to various AI systems. Currently, the QMS integrates a risk management system and a data management and governance system as sub-services.
Further sub-services, such as logging the use of the AI system (Art. 12 AIA), transparency approaches on how to create instructions for use (cf. Art. 13 AIA), and a human-machine interface for users of the AI system (cf. Art. 14 AIA), will be added in future versions of the prototype QMS. Future versions of the prototype QMS aim to integrate findings and technical features from earlier automated compliance verification research and insights from business process management research, such as modeling more extensive processes by combining regulations for AI development and post-development when the system is in use. Additionally, future work will further develop the various sub-services and conduct a broader evaluation to make the QMS even more comprehensive.

%
%
%
%

\end{document}